\documentclass[reprint,aps,prd,floats,floatfix,amsmath,amssymb,amsfonts,nofootinbib,longbibliography,superscriptaddress]{revtex4-2}
\usepackage[hyperindex,colorlinks]{hyperref}

\usepackage{graphicx}
\usepackage{dcolumn}
\usepackage{bm}

\begin{document}


\title{Dark Matter and Energy-Momentum Squared Gravity}

\author{H. R. Fazlollahi}
\email{seyed.hr.fazlollahi@gmail.com (Corresponding Author)}
\affiliation{%
 PPGCOSMO \& Departamento de Física, Universidade Federal do Espirito Santo (UFES), Av. Fernando Ferrari, 514 Campus de Goiabeiras, Vitória, Espírito Santo CEP 29075-910, Brazil}%

\author{H. Velten}%
 \email{hermano.velten@ufop.edu.br}
\affiliation{Departamento de Física, Universidade 
Federal de Ouro Preto, Campus Morro do Cruzeiro, Ouro Preto, MG, Brazil}

\author{H. Shojaie}%
 \email{h-shojaie@sbu.ac.ir}
\affiliation{Department of Physics, Shahid Beheshti University, G.C., Evin, Tehran 1983963113, Iran}

\begin{abstract}
In this study, we examine the energy-momentum squared modified theory of gravity, where the squared term $T_{\mu\nu}T^{\mu\nu}$ is
incorporated into the conventional gravitational Lagrangian. This modification aims to account for dark matter effects on
galactic scales. Specifically, we analyze the model near general relativity solutions for spherically symmetric and static metrics.
By fixing the components using the rotational velocities of galaxies, the model demonstrates the emergence of a flat rotation
curve in the galactic halo. Additionally, we investigate the proposed model's predictions for the light deflection angle and
radar echo delay. 
\begin{description}
\item[Keywords]
Energy-Momentum Squared Gravity; Dark Matter; Rotational Velocity.
\end{description}
\end{abstract}

\maketitle


\section{\label{sec:level1}INTRODUCTION}

One of the enduring puzzles in modern cosmology concerns the dark sector, encompassing dark matter and dark energy. The investigation of galactic rotation curves and the mass discrepancies in galaxy clusters led to the discovery of dark matter, which constitutes approximately $27\%$ of the matter-energy content of the Universe \cite{Bertone:2016nfn,Cirelli:2024ssz}. Conversely, the current accelerated expansion of the Universe is attributed to dark energy, which accounts for about $68\%$ of the Universe \cite{Planck:2018vyg}. Various proposals have been made to explain the dark sector \cite{Copeland:2006wr,Brax:2017idh,Oks:2021hef}. In order to keep agreement with observations and rely on the fundamental properties derived from particle physics and quantum mechanics, two notable dark energy models are the agegraphic \cite{Wei:2007ty,Neupane:2007ra,Fazlollahi:2022dwz,Zhang:2008mb} and holographic \cite{Li:2004rb,Oliveros:2012dua,Fazlollahi:2022kbv,Gong:2004fq} scenarios, which are based on the quantum properties of spacetime or the entropy of black holes.

 Axions like particles are, for example, interesting candidates, being a non-baryonic field proposed to account for dark matter properties \cite{Chadha-Day:2021szb,Duffy:2009ig,Panci:2022wlc}. Another prominent theory involves weakly interacting massive particles (WIMPs), a broad class of hypothesized particles that could contribute to dark matter \cite{Overduin:2004sz,Pospelov:2007mp,Roszkowski:2017nbc}. Heavy neutrinos have also been considered as potential dark matter candidates \cite{Dolgov:2000ew}. Despite these theoretical developments, both (in-)direct searches or accelerator experiments have not provided conclusive evidence for these particle-based scenarios \cite{Schumann:2019eaa}. Consequently, attention has shifted toward modified theories of gravity as an alternative approach to the particle dark matter paradigm \cite{Boehm:2024ana}. These theories incorporate additional terms into the standard Einstein-Hilbert action, allows dark matter to be interpreted as a geometric or matter-related evolution in spherically symmetric and static spacetimes \cite{Boehmer:2007kx,Camci:2021hpw}.  
 
 Similarly, dark energy can be modeled either as a result of purely geometric evolution \cite{Nojiri:2005vv,Fazlollahi:2018wmp,Artymowski:2015loa} or as a coupling between spacetime and matter in these modified gravities frameworks \cite{Myrzakulov:2012ug,Singh:2022eun,Nagpal:2018mpv}. Some theories also propose dark energy as non-Einsteinian matter fields \cite{Fazlollahi:2023cgp} or an effect of interior matter evolution \cite{Fazlollahi:2023rhg}. As such, modified theories of gravity offer powerful tools to unify the explanations of dark matter and dark energy within a single theoretical framework \cite{Nojiri:2008nt}.

In this study, we examine spherically symmetric and static spacetime to explore dark matter as an effect of a recently proposed modified theory, energy-momentum squared gravity. This theory extends the Einstein-Hilbert action by incorporating an additional term, $T_{\mu\nu}T^{\mu\nu}$ \cite{Roshan:2016mbt}. The term depends on the source matter field and introduces nonlinear density and pressure contributions to the field equations. As discussed in Ref. \cite{Roshan:2016mbt}, this alternative theory of gravity predicts a non-singular Universe. Furthermore, during the matter-dominated era, and given the small values of the theory's coupling constant, $\alpha$, the deviations from the standard $\Lambda$CDM model are negligible. This makes it compelling to investigate whether these additional terms can account for observations in galactic halos, eliminating the need to invoke mysterious masses as dark matter.

The paper is organized as follows: Section II provides a review of energy-momentum squared gravity. In Section III, we examine the field equations for spherically symmetric and static isolated systems in the vicinity of general relativity. Section IV explores the propagation of light within this typical galactic framework. Finally, the final remarks are presented in Section V.

Throughout this work, we adopt the geometrical unit system with $c=1$, while reducing Planck's mass as $m_{P}^2=\kappa^{-1}=(8\pi G)^{-1}$.

\section{ENERGY-MOMENTUM SQUARED GRAVITY REVIEW}

In this section, as a first step, we revisit energy-momentum squared gravity to derive the field equations. We also apply it the to static and spherically symmetric space-time in order to explore the basic correspondences between the metric functions and the source terms.

The action for this alternative theory is expressed as \cite{Roshan:2016mbt}:

\begin{equation}\label{eq:1}
    S=\frac{1}{2\kappa}\int\sqrt{-g}(R+\alpha T_{\mu\nu}T^{\mu\nu})\,d^4x+\int\sqrt{-g}L_{m}d^4x
\end{equation}
where $g$, $R$, and $\alpha$, respectively, refer to the determinant of the metric $g_{\mu\nu}$, the Ricci scalar, and the gravitational coupling constant of the modified gravity sector. The $L_{m}$ denotes the Lagrangian density, related to the energy-momentum tensor, i.e.,

\begin{equation}\label{eq:1}
    T_{\mu\nu}=-\frac{2}{\sqrt{-g}}\frac{\delta(\sqrt{-g}L_{m})}{\delta g^{\mu\nu}}.
\end{equation}
The variation of the action with respect to the inverse of the
metric tensor $g^{\mu\nu}$ obtains the field equations

\begin{equation}\label{eq:3}
    R_{\mu\nu}-\frac{1}{2}Rg_{\mu\nu}=\kappa T_{\mu\nu}^e
\end{equation}
where the effective energy-momentum tensor is defined as \cite{Fazlollahi:2023cgp}

\begin{eqnarray}\label{eq:4}
T_{\mu\nu}^e=T_{\mu\nu}+\frac{\alpha}{2}\Big[\frac{1}{2}T_{\alpha\beta}T^{\alpha\beta}g_{\mu\nu}+2L_{m}(T_{\mu\nu}-\frac{1}{2}Tg_{\mu\nu}) \nonumber \\  +TT_{\mu\nu}-2T_{\mu}^{\alpha}T_{\nu\alpha}+4T^{\alpha\beta}\frac{\partial^2 L_{m}}{\partial g^{\mu\nu}\partial g^{\alpha\beta}}\Big].
\end{eqnarray}

With this strategy the field equation \eqref{eq:3} resembles the GR structure in the sense it keeps all $T_{\mu\nu}$ dependence in the right hand side of the field equations.

By considering the energy-momentum tensor assumes the perfect fluid form

\begin{equation}\label{eq:5}
    T_{\mu\nu}=(\rho+p)u_{\mu}u_{\nu}+pg_{\mu\nu}
\end{equation}
in which $\rho$, $p$ and $u_{\mu}$, respectively are the energy density, pressure, and the comoving four-velocity satisfying the conditions $u_{\mu}u^{\mu}=-1$, we obtain,

\begin{equation}\label{eq:6}
    T=T_{\mu}^{\mu}=-\rho+3p\quad {\rm and} \quad T_{\mu\nu}T^{\mu\nu}=\rho^2+3p^2.
\end{equation}
Now by taking Lagrangian $L_{m}=-\rho$, and using relations \eqref{eq:6} in \eqref{eq:4}, the modified Einstein’s field equations \eqref{eq:3} read as

\begin{multline}\label{eq:8}
    R_{\nu}^{\mu}-\frac{1}{2}R\delta_{\nu}^{\mu}=\kappa((\rho+p)u_{\nu}^{\mu}+p\delta_{\nu}^{\mu})-\alpha\Big(\frac{1}{2}(\rho^2-9p^2)\delta_{\nu}^{\mu}+\\3(\rho^2- p^2)u_{\nu}u^{\mu}+2(\rho^2+3p^2)-4g^{\mu\mu}T^{\alpha\beta}\frac{\partial^2 \rho}{\partial g^{\mu\nu}\partial g^{\alpha\beta}}\Big)
\end{multline}
The right-hand side reveals that we are no longer dealing with a standard perfect fluid but rather with an effective fluid. The cosmological applications of this theory have been explored across various dark energy scenarios (e.g., \cite{Ranjit:2020syg},\cite{Akarsu:2023nyl},\cite{Board:2017ign}). Additionally, studies in recent years have considered gravastars and compact stars within the framework of this theory \cite{Nari:2018aqs},\cite{Sharif:2022dzl},\cite{Sharif:2023uac}.

In this work, however, we employ this alternative theory to investigate dark matter in an isolated system described by a static, spherically symmetric metric.

\begin{equation}\label{eq:8}
    ds^2=-e^{a(r)}dt^2+e^{b(r)}dr^2+r^2d\theta^2+r^2sin^2\theta d\phi^2
\end{equation}
To maintain simplicity, the effective energy-momentum tensor of the matter is modeled as a pressureless fluid, where the $L_{m}=p$ is assumed to depend solely on the metric components without their derivatives. Consequently, the various components of the field equations can be expressed as follows,

\begin{equation}\label{eq:9}
    G_{t}^t=\frac{1}{r^2e^b}-\frac{b'}{re^b}-\frac{1}{r^2}=-\kappa\rho+\frac{\alpha}{2}\rho^2
\end{equation}
\begin{equation}\label{eq:10}
    G_{r}^r=\frac{1}{r^2e^b}+\frac{a'}{re^b}-\frac{1}{r^2}=-\frac{5\alpha}{2}\rho^2
\end{equation}
\begin{equation}\label{eq:11}
    G_{\theta}^\theta=G_{\phi}^\phi=\frac{1}{4e^b}(2a''+a'^2-a'b')+\frac{a'-b'}{2re^b}=-\frac{5\alpha}{2}\rho^2,
\end{equation}
where the prime $'$ denotes differentiation with respect to the radial coordinate $r$. As shown, in the absence of pressure, the energy-momentum squared term introduces equal radial and tangential pressures into the model, proportional to the square of the matter density $\rho^2$. Due to the presence of $\rho^2$ in equations \eqref{eq:9}-\eqref{eq:11}, obtaining exact solutions for the metric components in expression \eqref{eq:8} is not straightforward. To address this, one can eliminate $\rho^2$ by subtracting equation \eqref{eq:10} from equation \eqref{eq:9}, as demonstrated below.

\begin{equation}\label{eq:12}
    \frac{a'+b'}{re^b}=\rho(\kappa-3\alpha\rho)
\end{equation}
This relation provides a linear combination of the derivatives of the metric functions in terms of both the first and second powers of the density $\rho$. It proves useful in the following section for determining the exact forms of $a$, $b$, and $\rho$.

\section{GALACTIC DARK MATTER FEATURES}

To determine the set of functions ${a, b, \rho}$ in this model, it is essential to establish a relationship between the time and radial components. Due to the influence of the matter field, deviations in spacetime around a central massive object are expected. In this context, if the combination $a'+b'$ forms a well-defined differential expression, a solution of the form 
\begin{equation}\label{eq:13}
    e^{a(r)+b(r)}=l(r),
\end{equation}
can be obtained as in \cite{Sobouti:2006rd}. In this proposal, one expects the function $l(r)$ to be slightly different from $1$. As a plausible form, we can assume

\begin{equation}\label{eq:14}
    l(r)=\left(\frac{r}{\sigma}\right)^s,
\end{equation}
where $s$ is a dimensionless parameter and $\sigma$ is the length scale of the system. To remain in the vicinity of general relativity, it is necessary to have

\begin{equation}\label{eq:15}
    s\ll 1,
\end{equation}
that gets $l(r)\approx 1+s\ln(r/\sigma)$. Hence, and with an eye to \eqref{eq:13}, one obtains

\begin{equation}\label{eq:126}
    a'+b'=\frac{s}{r},
\end{equation}
Plugging the above relation into \eqref{eq:12} and using \eqref{eq:13} gives metric components as a function of density,

\begin{equation}\label{eq:17}
    e^b=\frac{s}{r^2\rho(\kappa-3\alpha\rho)},
\end{equation}

\begin{equation}\label{eq:18}
    e^a=\frac{\rho r^2(\kappa-3\alpha\rho)}{s}\left(\frac{r}{\sigma}\right)^s,
\end{equation}
that leads to the following relations

\begin{equation}\label{eq:19}
    \frac{r\rho'(\kappa-6\alpha\rho)}{s}+\frac{3\rho(\kappa-3\alpha\rho)}{s}-\frac{1}{r^2}=-\kappa\rho+\frac{\alpha}{2}\rho^2,
\end{equation}

\begin{equation}\label{eq:20}
    \frac{r\rho'(\kappa-6\alpha\rho)}{s}+\frac{3\rho(\kappa-3\alpha\rho)(3+s)}{s}-\frac{1}{r^2}=-\frac{5\alpha}{2}\rho^2,
\end{equation}

\begin{multline}\label{eq:21}
    \frac{r^2\rho''(\kappa-6\alpha\rho)}{2s}-\frac{3\alpha r^2\rho'^2}{s}+\frac{3r\rho'(\kappa-6\alpha\rho)(4+s)}{4s}+\\\frac{\rho(\kappa-3\alpha\rho)(s^2+6s+12)}{4s}=-\frac{5\alpha}{2}\rho^2,
\end{multline}
By eliminating $\rho^2$ from the left-hand side of equation \eqref{eq:19}, we arrive at:

\begin{multline}\label{eq:22}
    \frac{r^2\rho''(\kappa-6\alpha\rho)}{2s}-\frac{3\alpha r^2\rho'^2}{s}+\frac{r\rho'(\kappa-6\alpha\rho)(8+3s)}{4s}\\-\frac{3\alpha(2+s)\rho^2}{4}+\frac{\kappa\rho(2+s)}{4}+\frac{1}{r^2}=0.
\end{multline}
The above equation allow us to obtain two different solutions for the density function as a function of $r$. They read,

\begin{equation}\label{eq:23}
    \rho_{1}=\frac{\kappa}{6\alpha}-\frac{1}{6\alpha r}\sqrt\frac{\kappa^2r^2s^2+4s(12\alpha-\kappa^2r^2)-4\kappa^2r^2}{s^2-4s-4}
\end{equation}

\begin{equation}\tiny\label{eq:25}
    \rho_{2}=\frac{\kappa}{6\alpha}+\sqrt{\frac{1}{6\alpha r}\frac{(18+s)(2+s)(\kappa^2r^2s^2+4s(12\alpha-\kappa^2r^2)-4\kappa^2r^2)}{(s^2+20s+36)(s^2-4s-4)}}
\end{equation}
Although solving equation \eqref{eq:22} yields two distinct density profiles, for $s\ll1$, both solutions converge to:

\begin{equation}\label{eq:25}
    \rho_{1}=\rho_{2}\approx\frac{s}{\kappa r^2}-\frac{(\kappa^2 r^2-3\alpha)}{\kappa^3r^4}s^2+O(s^3).
\end{equation}
Investigating \eqref{eq:25} shows that the density rapidly decreases
with radius proportionally with $\rho\propto r^{-2}$.

Using the above density in \eqref{eq:17} and \eqref{eq:18}, we find

\begin{equation}\label{eq:26}
    e^a=(1-s)\left(\frac{r}{\sigma}\right)^s+\frac{3\alpha s^2}{\sigma^s\kappa^2}\left(\frac{2-s}{r^{2-s}}\right)-\frac{18\alpha^2s^2}{\sigma^s\kappa^4}\left(\frac{1-s}{r^{4-s}}\right),
\end{equation}

\begin{equation}\label{eq:27}
    e^b=1-s+\frac{3\alpha s^2}{\kappa^2}\left(\frac{2-s}{r^2}\right)-\frac{18\alpha^2s^2}{\kappa^4}\left(\frac{1-s}{r^4}\right),
\end{equation}
which for $s\ll1$ are, approximately

\begin{multline}\label{eq:28}
    e^a\approx1-s+s\ln\left(\frac{r}{\sigma}\right)+\left(\frac{6\alpha}{\kappa^2r^2}-\frac{18\alpha^2}{\kappa^4r^4}\right)s^2-\\\ln\left(\frac{r}{\sigma}\right)\left(1-\frac{1}{2}\ln\left(\frac{r}{\sigma}\right)\right)s^2,
\end{multline}

\begin{equation}\label{eq:29}
    e^b\approx1+s+\left(1-\frac{6\alpha}{\kappa^2r^2}+\frac{18\alpha^2}{\kappa^4r^4}\right)s^2.
\end{equation}
From the above solutions one can infer that the model is flat, asymptotically.
To investigate whether such a model represents dark matter as an effect of the energy-momentum squared term, we consider a test particle moving along a time-like geodesic orbit in the plane $\theta=\pi/2$. The corresponding geodesic equation for the $r$-coordinate is given by \cite{Weinberg:1972kfs}:

\begin{equation}\label{eq:30}
    \frac{d^2r}{d\tau^2}+\frac{a'e^{a(r)}}{2e^{b(r)}}\left(\frac{dr}{d\tau}\right)^2+\frac{b'}{2}\left(\frac{dr}{d\tau}\right)^2-\frac{r}{e^{b(r)}}\left(\frac{d\phi}{d\tau}\right)^2=0
\end{equation}

Here, $\tau$ denotes the affine parameter along the geodesic. The momenta $P_{0}$ and $P_{3}$ associated with the test particle's motion along its geodesics in a spherically symmetric and static spacetime are conserved \cite{Weinberg:1972kfs}, i.e.,

\begin{equation}\label{eq:31}
    \mathcal{E}=e^{a(r)}\left(\frac{dt}{d\tau}\right)=const,
\end{equation}

\begin{equation}\label{eq:32}
    \mathcal{J}=r^2\left(\frac{d\phi}{d\tau}\right)^2=const,
\end{equation}
where $\mathcal{E}$ and $\mathcal{J}$ are, respectively, the energy and the $\phi$-coordinate of the angular momentum of the test particle. For motion on the stable circular orbits, i.e. $dr/d\tau=0$, the geodesic equation \eqref{eq:30} can be written as

\begin{equation}\label{eq:33}
    \frac{a'\mathcal{E}^2}{2e^{a(r)}}=\frac{\mathcal{J}^2}{r^3}.
\end{equation}
Far away from the gravitational source, as a weak-field approximation, the circular orbital speed becomes \cite{Lidsey:2001nj}

\begin{equation}\label{eq:34}
    \upsilon=\frac{r}{e^{a(r)/2}}\left(\frac{d\phi}{dt}\right).
\end{equation}
With the aid of \eqref{eq:32} and \eqref{eq:33}, the above orbital speed can be written in terms of the derivative of the metric function. Its square value reads

\begin{equation}\label{eq:35}
    \upsilon^2=\frac{ra'}{2}.
\end{equation}

The observation of flatness in galactic rotation curves suggests that the function in the halo of galaxies is independent of the distance $r$. This phenomenon is recognized as one of the strongest indicators of the existence of dark matter. By using the density expression \eqref{eq:25} and calculating the square of the orbital speed \eqref{eq:35} for the case $s\ll1$,

\begin{equation}\label{eq:36}
    \upsilon^2\approx\frac{s}{2}-\frac{6\alpha}{\kappa^4r^4}(\kappa^2r^2-6\alpha)s^2.
\end{equation}
It is demonstrated that the tangential speed of the test particle is approximately independent of the radial coordinate in the galactic halo. Consequently, the model, through the role of the energy-momentum squared term in the field equations for pressureless matter, effectively illustrates the presence of dark matter.
The tangential velocity in stable orbits around the galactic halo typically ranges from $200$ to $500 km/s$ \cite{Binney:1993ce}. For a typical spiral galaxy, this corresponds to $\upsilon^2\approx O(10^{-6})$ in $c^2$ units. Consequently, by neglecting higher-order terms of $s$, equation  \eqref{eq:36} reveals that $s/2$ is approximately equal to the square of the tangential velocity, i.e., $s/2\approx\upsilon^2$. Although this assumption suggests that the coupling parameter $\alpha$ has no significant influence on $\upsilon$, $\alpha$ remains a critical parameter for constructing a valid model that aligns with observations in the galactic halo.

\section{DARK MATTER SIGNATURES}

In this section, we explore dark matter signatures resulting from parameterization \eqref{eq:14} by examining light-deflection angles and radio echo delay, as detailed in the following subsections.

\subsection{Light-Deflection Angles}

One method to detect the effects of dark matter is through the study of gravitational lensing around and within the halos of galaxies. Specifically, we analyze the light deflection angles, where $\Delta\phi$ is defined as follows:

\begin{equation}\label{eq:37}
    \Delta\phi=2|\phi(r_{0})-\phi(\infty)|-\pi,
\end{equation}
here $r_{0}$ denotes the region of flat rotation curves. Using geodesic equation \eqref{eq:30} for a photon yields \cite{Weinberg:1972kfs}

\begin{equation}\label{eq:38}
    \phi(r_{0})-\phi(\infty)=\int_{r_{0}}^\infty\left(\frac{e^{b(r)}}{{e^{a(r_{0})-a(r)}\left(\frac{r}{r_{0}}\right)^2}}\right)^{(1/2)}\frac{dr}{r}.
\end{equation}
Since $s\ll1$, only by keeping the first two terms of metric functions \eqref{eq:26} and \eqref{eq:27}, we find

\begin{multline}\label{eq:39}
    \phi(r_{0})-\phi(\infty)=\sqrt{1+s}\int_{r_{0}}^{r_{d}}\left(\left(\frac{r}{r_0}\right)^{2-s}-1\right)^{-1/2}\frac{dr}{r}+\\\int_{r_d}^{\infty}\left(\frac{1-\frac{2GM}{r_0}}{1-\frac{2GM}{r}}\left(\frac{r}{r_0}\right)^2-1\right)^{-1/2}\frac{dr}{r}.
\end{multline}
Here, $r_d$ represents the radius beyond which the contribution of dark matter diminishes. Consequently, the first integral corresponds to the dark matter-dominated region, while the second integral describes the exterior region of the dark matter halo, governed by the Schwarzschild metric. By applying the Robertson expansion \cite{Weinberg:1972kfs}, we obtain:

\begin{multline}\label{eq:40}
    \Delta\phi+\pi=2|\frac{2\sqrt{1+s}}{2-s}\arctan\left(\sqrt{\left(\frac{r}{r_0}\right)^{2-s}-1}\right)+\\\arcsin\left(\frac{r_0}{r_d}\right)+\frac{GM}{r_0}\left[2-\sqrt{1-\left(\frac{r_0}{r_d}\right)^2}-\sqrt{\frac{r_d-r_0}{r_d+r_0}}\right]|.
\end{multline}
It is evident that as $r_d=r_0$, the effects of dark matter disappear, and the deflection angle is solely due to baryonic matter. In this specific case, one obtains $\Delta\phi=4GM/r_0$, which aligns with the predictions of general relativity \cite{Weinberg:1972kfs}. To constrain the model, it is necessary to utilize observational data, including the approximate mass and halo radius of galaxies. In Figure 1, the effects of dark matter on the light deflection angle $\Delta\phi$ for $s=10^{-6}$, $2\times10^{-6}$, and $3\times 10^{-6}$, as well as for four different galaxies listed in Table 1, are illustrated. As shown, increasing the value of $s$ results in a larger light deflection angle. Thus, the parameter $s$ can be interpreted as a representation of the effects of dark matter on the geometry of spacetime.

\begin{table}[ht]
\centering
\begin{tabular}{c c c}
\hline\hline
Galaxy & $r_d(kpc)$ & M($10^{10} M_\odot$) \\ [0.5ex]
\hline
NGC 5533 & 72.0 & 22.0 \\
NGC 3992 & 30.0 & 16.22 \\
NGC 5907 & 32.0 & 10.8 \\
NGC 2998 & 48.0 & 11.3 \\ [1ex]
\hline
\end{tabular}
\caption{Data used in this work for four different galaxies \cite{Sanders:2002pf}.}
\label{table:nonlin}
\end{table}

\begin{figure}[ht!]
    \centering
    \includegraphics[width=65mm]{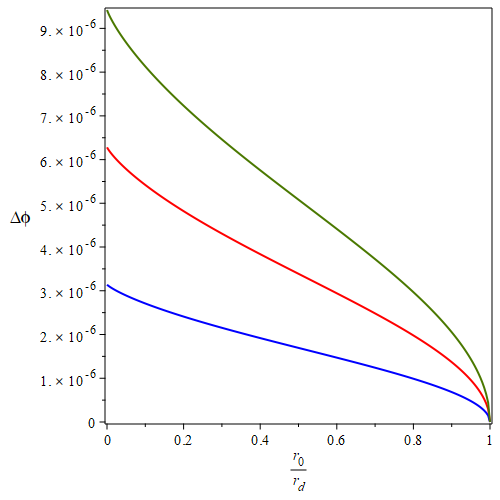}
    \caption{The light-deflection angle for different galaxies, given in Table 1, for $s=10^{-6}$ (blue), $2 \times 10^{-6}$ (red) and $3 \times 10^{-6}$ (green).}
    \label{fig1}
\end{figure}

\subsection{Radar Echo Delay}

One of the key consequences of relativity theory is time dilation, which occurs when light travels near massive bodies. This phenomenon can be observed through the measurement of radar echo delays. Shapiro proposed a method to detect this effect, now recognized as the fourth test of general relativity \cite{Shapiro:1964uw}. For photons propagating from $r=r_1$ to $r=r_2$, with $\theta=\pi/2$, the elapsed time is expressed as \cite{Weinberg:1972kfs}:

\begin{equation}\label{eq:41}
    t(r_1,r_2)=\int_{r_1}^{r_2}\frac{e^{b(r)/2}}{e^{a(r)/2}}\left(1-\frac{e^{a(r)}}{e^{a(r_1)}}\left(\frac{r_1}{r}\right)^2\right)^{-1/2}dr.
\end{equation}

By setting $r_1=r_0$ in our model, where $r_2$ represents a distant point far from the region surrounding the galaxy, we can split the integral above into two parts:

\begin{multline}\label{eq:42}
    t(r_1,r_2)=\int_{r_0}^{r_d}\frac{e^{b(r)/2}}{e^{a(r)/2}}\left(1-\frac{e^{a(r)}}{e^{a(r_1)}}\left(\frac{r_1}{r}\right)^2\right)^{-1/2}dr+\\\int_{r_d}^{r}\frac{r}{r-2GM}\left(1-\frac{1-\frac{2GM}{r}}{1-\frac{2GM}{r_d}}\left(\frac{r_d}{r}\right)^2\right)^{-1/2}dr,
\end{multline}
which yields,

\begin{multline}\label{eq:43}
    t(r_0,r)=\frac{2}{(1-s)(2-s)}\sqrt{\frac{\sigma^2}{r_d^{s-2}}}\sqrt{1-\left(\frac{r_d}{r_0}\right)^{s-2}}+\\\sqrt{r^2-r_d^2}+2GM\ln\left(\frac{r+\sqrt{r^2-r_d^2}}{r_d}\right)+GM\left(\frac{r-r_d}{r+r_d}\right)^{1/2}.
\end{multline}

The first term represents the effect of the region with flat rotation curves. In the absence of this region, when  $r_d=r_0$, we expect the time delay predicted by general relativity, which occurs in the absence of dark matter. Analysis of the model shows that the free parameter $\sigma$ plays no significant role when we assume $\sigma\ll r_d$. To better illustrate the effects of dark matter on radar echo delay, it is useful to investigate Eq. \eqref{eq:43} by fixing the mass and varying $r_d$. In this context, the variation of radar echo delay Eq. \eqref{eq:43} is plotted for $M=2.2\times10^{11} M_\odot$ (a reasonable value for the galactic mass), $r=10r_d$ and $\sigma=1$, for different values of $r_d=72$, $70$ and $60$ kpc. As shown, increasing the value of $r_d$ (which corresponds to more dark matter) leads to an increase in the radar echo delay.

\begin{figure}[ht!]
    \centering
    \includegraphics[width=65mm]{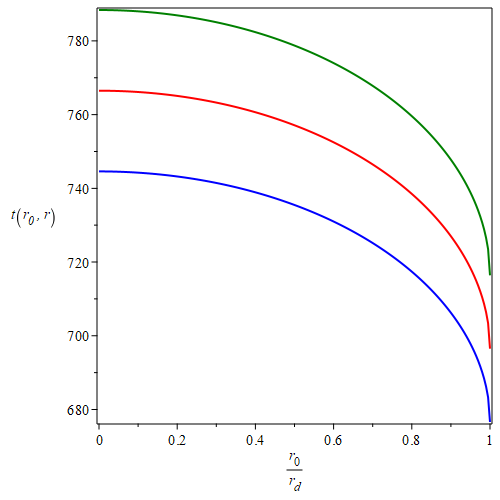}
    \caption{The radar echo delay as a function of $r_0/r_d$ when $M=M=2.2\times10^{11} M_\odot$, $r=10r_d$, $\sigma=1$, and $s=10^{-6}$, when $r_d=72$ (green curve), $70$ (red curve), and $68$ (blue curve).}
    \label{fig1}
\end{figure}

\section{REMARKS}

Dark matter, one of the most enigmatic aspects of astrophysics, challenges the conventional framework of general relativity. With advancements in high-energy physics and particle physics, efforts have been made to interpret dark matter as a form of baryonic or non-baryonic particles. However, these particle-based models have yet to gain substantial support from observational data in particle accelerators.
An alternative approach involves exploring modified theories of gravity as a promising framework, wherein dark matter is interpreted as a geometric-matter coupling effect. In this work, we investigated the recently proposed "energy-momentum squared gravity" theory in this context. As discussed, analyzing a model for a dust field in a spherically symmetric and static spacetime close to the framework of general relativity reveals that dark matter can be interpreted as an effect of the squared matter density. Actually, the modified gravity features appears due to the adoption of relation \eqref{eq:14} since the parameter $\alpha$ enters into the dynamics as a second order effect as seen in \eqref{eq:36}. Additionally, through the study of light deflection angles and radar echo delays, we demonstrate that the model effectively accounts for the presence of significant dark matter in the outer regions of galactic halos.
It is important to note that this study focuses on dark
matter under weak-field conditions, where the gravitational
effects of the central object are minimal. Moreover, we
constrained our analysis to non-relativistic fluids. Future
investigations should extend this work to explore models
involving strong gravity or relativistic fluids, which could
provide deeper insights into the nature of dark matter.

\acknowledgments{HF \& HV thank the Research Council of UFES for financial support. HV thanks FAPEMIG and CNPq for financial support. HF also thanks A.H. Fazlollahi for suggesting the consideration of this model.}

Data Availability Statement: No Data associated in the manuscript

\end{document}